\documentclass[aps,prl,reprint,showpacs,superscriptaddress]{revtex4-1}
\usepackage{color}
\usepackage{amssymb}
\usepackage{amsmath}
\usepackage{graphicx}

\begin{document}

\setcounter{page}{1}

\title{A New Phosphorus Allotrope with Direct Band Gap and High Mobility}

\author{W. H. Han}
\affiliation{Department of Physics, Korea Advanced Institute of Science and Technology, Daejeon 34141, Korea}
\author{Sunghyun Kim}
\affiliation{Department of Physics, Korea Advanced Institute of Science and Technology, Daejeon 34141, Korea}
\author{In-Ho Lee}
\affiliation{Korea Research Institute of Standards and Science, Daejeon 34113, Korea}
\author{K. J. Chang}
\email[]{kjchang@kaist.ac.kr}
\affiliation{Department of Physics, Korea Advanced Institute of Science and Technology, Daejeon 34141, Korea}

\bibliographystyle{apsrev4-1}

\date{\today}
\begin{abstract}
Based on \emph{ab initio} evolutionary crystal structure search computation,
we report a new phase of phosphorus called green phosphorus ($\lambda$-P), which exhibits the direct band gaps ranging from 0.7 to 2.4 eV and the strong anisotropy in optical and transport properties.
Free energy calculations show that a single-layer form, termed green phosphorene, is energetically more stable than blue phosphorene and a phase transition from black to green phosphorene can occur at temperatures above 87 K.
Due to its buckled structure, green phosphorene can be synthesized on corrugated metal surfaces rather than clean surfaces.
\end{abstract}


\maketitle

The successful isolation of graphene \cite{Novoselov2004}, a single layer of carbon atoms in a two-dimensional (2D) honeycomb lattice, has generated tremendous interest in 2D layered materials \cite{Geim2013}.
Various applications using graphene have been explored based on the unusual properties such as massless Dirac fermions, high mobility, and high thermal conductivity \cite{Novoselov2005,Zhang2005}.
However, the gapless nature of graphene has been an obstacle in practical applications for electronic devices due to its low on/off ratios \cite{Yang2012}.
Transition metal dichalcogenides belonging to the family of 2D materials have the semiconducting gaps in the visible range, but the carrier mobility in thin films is significanlty reduced \cite{Ling2015}.
Recently, atomically thin black phosphorus has been successfully separated from its layered bulk \cite{Liu2014,Li2014}, and this emerging 2D material with the tunable band gap by varying the number of atomic layers bridges the gap between graphene and transition metal dichalcogenides \cite{Tran2014,Li2016}.
Due to its high anisotropic mobility, black phosphorus is considered as a promising material for electronic and optoelectronic devices \cite{Qiao2014,Liu2014,Li2014,Wang2014,Ling2015}.

Elemental phosphorus is known to exist in several three-dimensional (3D) allotropes, such as red, white, and violet phosphorus, besides black phosphorus ($\alpha$-P) which is the most stable phase among them.
Due to the presence of various phases, it is expected that an unknown phosphorus allotrope will form under the control of substrate, temperature, and pressure.
Zhu and Tom$\mathrm{\acute{a}}$nek proposed a new stable phase of phosphorus called blue phosphorus ($\beta$-P), with the structural similarity to a buckled graphene \cite{Zhu2014}.
Unlike black phosphorus, blue phosphorus displays the indirect band gaps, regardless of the number of atomic layers.
Other 2D structures such as $\gamma$-, $\delta$-, $\varepsilon$-, $\zeta$-, $\eta$-, $\theta$-, and $\psi$-phosphorene were later suggested based on theoretical calculations \cite{Guan2014,Zhao2015,Wu2015,Schusteritsch2016,Wang2016}.
Recently, blue phosphorene has been successfully synthesized on the Au(111) substrate by molecular beam epitaxy \cite{Zhang2016}.
The realization of blue phosphorene not only opens up the potential of various metastable allotropes but also motivates research into a new level of phosphorus that provides exciting characteristics such as broad band gaps and high anisotropic mobility.

In this work, we use an {\it ab initio} evolutionary crystal structure search method to explore a new phosphorus allotrope called green phosphorus. 
The new P allotrope belongs to a class of 2D materials and has the direct band gap characteristics in the range of 0.7$-$2.4 eV, depending on the number of atomic layers.
Because of the weak interlayer interaction, green phosphorus should be easily exfoliated to form thin films.
A monolayer of green phosphorus exhibits the high in-plane anisotropy in optical absorption and transport properties, suitable for applications in nanoelectronics and nanophotonics.
We discuss the transition pathway from black to green phosphorene and the effects of temperature and substrate on the synthesis of green phosphorene.

The structural optimization and electronic structure calculations were performed using the functional form proposed by Perdew, Burke, and Ernzerhof (PBE) \cite{Perdew1996} for the exchange-correlation potential within the framework of density functional theory and the projector augmented wave potentials \cite{Blochl1994}, as implemented in the VASP code \cite{Kresse1996}.
To describe more accurately interlayer interactions, we additionally used the PBE+D2 functional which includes van der Waals forces \cite{Grimme2006}, yielding the lattice constants (Table 1) in good agreement with the experimentally measured values for black phosphorus \cite{Brown1965}.
The wave functions were expanded in plane waves up to an energy cutoff of 500 eV.
For Brillouin zone (BZ) integration, we used the $k$-point sets generated by the $16 \times 16 \times 1$ and $12 \times 12 \times 12$ Monkhorst-Pack meshes for 2D and 3D systems, respectively.
The ionic coordinates were fully optimized until the residual forces were less than 0.01 eV/$\mathrm{\AA}$.

\begin{figure*}[!t]
\includegraphics[width=2\columnwidth]{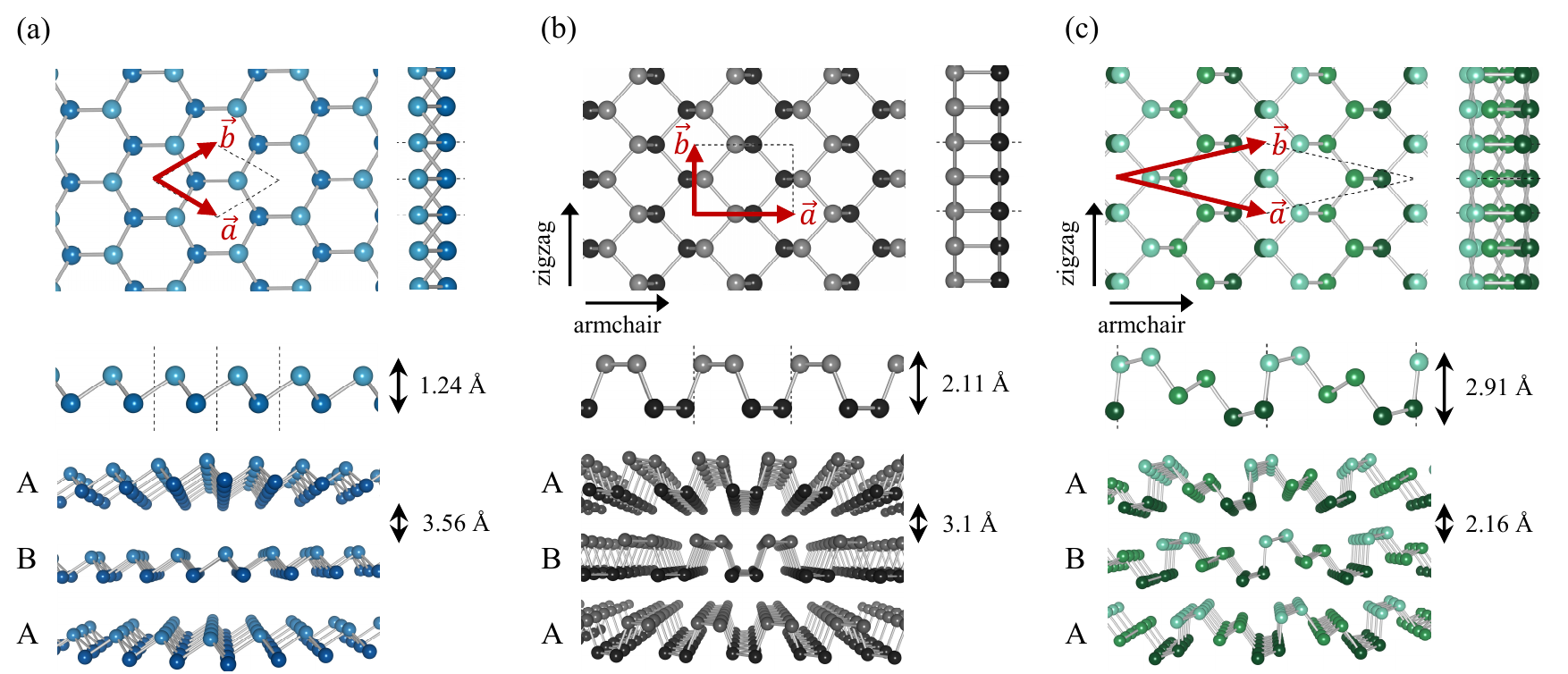}
\caption{\label{fig:1} Atomic structures for (a) $\beta$-P (blue), (b) $\alpha$-P (black), and (c) $\lambda$-P (green). 
Top and side views of a monolayer in upper panels whereas $AB$-stacked atomic layers in lower panels. }
\end{figure*}

For phosphorus systems with $N$ atoms per unit cell ($N = 2k$, $3 \leq k \leq 12$), we explored low-energy allotropes by using an {\it ab initio} evolutionary crystal structure search method, as implemented in the AMADEUS code \cite{Lee2016}.
Distinct configurations were generated by employing the conformational space annealing algorithm for global optimization.
We find that black phosphorus is the most stable allotrope at zero temperature, as reported previously \cite{Bachhuber2014}.
Among various metastable configurations \cite{*[{See Supplemental Material at }] [{}] supp}, we obtained a novel allotrope in the space group $C2/m$, which contains 6 atoms in the primitive cell.
The new metastable allotrope, termed green phosphorus ($\lambda$-P), has a layered structure like black and blue phosphorus (Fig. 1).

\begin{table*}[!t]\centering
\begin{ruledtabular}
\caption{The in-plane lattice parameters ($a$ and $b$), interlayer spacings ($d$), layer thicknesses ($t$), relative energies (${\Delta}E$), and interlayer interaction energies ($E_{int}$) are compared for $\beta$-P (blue), $\alpha$-P (black), and $\lambda$-P (green). The PBE-D2 and PBE functionals for the exchange-correlation potential are used for bulk and monolayer structures, respectively. Numbers in parentheses are the experimentally measured values for black phosphorus \cite{Brown1965}. }
\begin{tabular}{cccccccccc}
     &  \multicolumn{5}{c}{Bulk} & \multicolumn{4}{c}{Monolayer}\\
\cline{2-6} \cline{7-10}
Allotrope    & $a$ ($\mathrm{\AA}$) & $b$ ($\mathrm{\AA}$) & $d$ ($\mathrm{\AA}$)  & ${\Delta}E$ (meV/atom) & $E_{int}$ (meV/atom) & $a$ ($\mathrm{\AA}$) & $b$ ($\mathrm{\AA}$) &  $t$ ($\mathrm{\AA}$)  & ${\Delta}E$ (meV/atom)  \\
\hline
Blue P  & 3.29 & 3.29 & 3.56 &  85.7 & 32.6 & 3.28 & 3.28 & 1.24 & 1.8  \\
Black P & 4.44 & 3.32 & 3.10 &   0   & 79.2 & 4.62 & 3.30 & 2.11 &  0   \\
        & (4.38) & (3.31) & (3.05) &  &  &  &  &  &  \\
Green P & 7.13 & 7.13 & 2.16 &  19.0 & 73.9 & 7.27 & 7.27 & 2.91 & 0.9  \\

\end{tabular}
\end{ruledtabular}
\end{table*}

Figure 1 illustrates the structural relationship between black, blue, and green phosphorus.
Black and blue phosphorus exhibit distinct structural differences in side view, such as armchair and zigzag ridges, respectively.
Because the zigzag ridges are less puckered, the layer thickness of blue phosphorus is reduced to 1.24 $\mathrm{\AA}$, compared to black phosphorus (2.11 $\mathrm{\AA}$).
In black phosphorene, reversing all bonds in every fourth row from an up-position to a down-position (or vice versa) converts all the armchair ridges to zigzag ridges, leading to the transformation to blue phosphorene \cite{Zhu2014}.
In green phosphorus, the $AB$ stacking of atomic layers is energetically more favorable by 15 meV/atom than the $AA$ stacking.
All the atoms are three-fold coordinated, with the bond lengths of 2.23 and 2.26 $\mathrm{\AA}$ for the in-plane and out-of-plane bonds, respectively.
Since a monolayer of green phosphorus, termed green phosphorene, contains three slightly buckled atomic layers, it has the largest layer thickness of 2.91 $\mathrm{\AA}$ among the three allotropes.
Green phosphorene consists of the armchair and zigzag ridges, thus, its structure is characterized by the combination of black and blue phosphorene (Fig. 2). 
The transformation from black to green phosphorene can be viewed by flipping every twelfth row of bonds from the up-position to down-position by dislocations after every fourth row of the armchair ridges are reversed to the opposite side of the 2D plane.

\begin{figure}[h]
\includegraphics[width=\columnwidth]{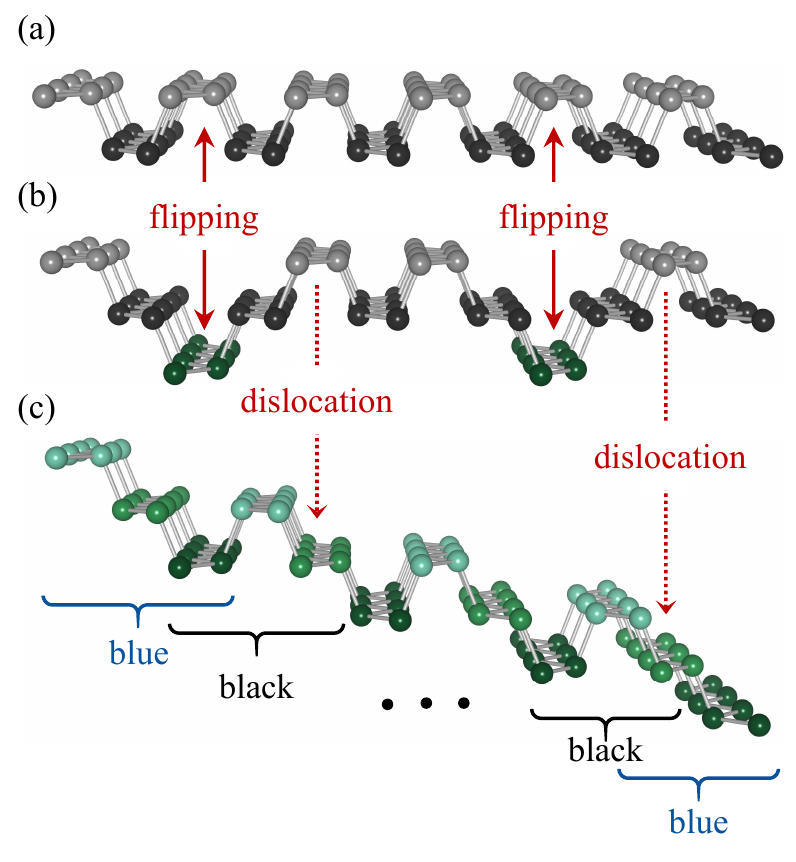}
\caption{\label{fig:2} Schematic view of the conversion of (a) $\alpha$-P (black) to (c) $\lambda$-P (green) via (b) an intermediate structure formed by flipping every fourth row of the armchair ridges. } 
\end{figure}

The lattice parameters and energetics of black, blue, and green phosphorus are compared in Table 1.
Green phosphorus is more stable by 67 meV/atom than blue phosphorus, while its energy is higher by 19 meV/atom than that of black phosphorus.
Although green phosphorus has the smallest interlayer distance of $d$ = 2.16 $\mathrm{\AA}$, the interlayer interaction of 74 meV/atom is comparable to that of black phosphorus (79 meV/atom), indicating that individual layers can be mechanically exfoliated. 
As the number of atomic layers is reduced to a single layer, the in-plane lattice parameters slightly increase from $a$ = $b$ = 7.13 $\mathrm{\AA}$ to $a$ = $b$ = 7.27 $\mathrm{\AA}$.
However, the relative stability of the three P allotropes is not affected by forming an isolated monolayer.
The energies of green and blue phosphorene relative to black phosphorene are 0.9 and 1.8 meV/atom, respectively.
Because these energy differences are extremely small, the energetics of black, blue, and green phosphorene are sensitive to external parameters such as substrate and temperature to be discussed later.

To determine whether green phosphorus maintains its crystal structure under ambient conditions,
we examined the phonon spectra of bulk and monolayer green phosphorus and found no imaginary phonon modes over the entire BZ \cite{*[{See Supplemental Material at }] [{}] supp}, indicating that both the structures are dynamically stable.
Similar to blue phosphorus, green phosphorus exhibits more rigid longitudinal optical modes than black phosphorus.
The thermal stability of bulk and monolayer green phosphorus was verified by performing \emph{ab initio} molecular dynamics simulations up to 100 ps at high temperatures of $600-800$ K \cite{*[{See Supplemental Material at }] [{}] supp}.
We also calculated the elastic constants and confirmed that stress-strain relations meet the criteria for mechanical stability in both the bulk and monolayer structures \cite{*[{See Supplemental Material at }] [{}] supp}.

\begin{figure}[!h]
\includegraphics[width=\columnwidth]{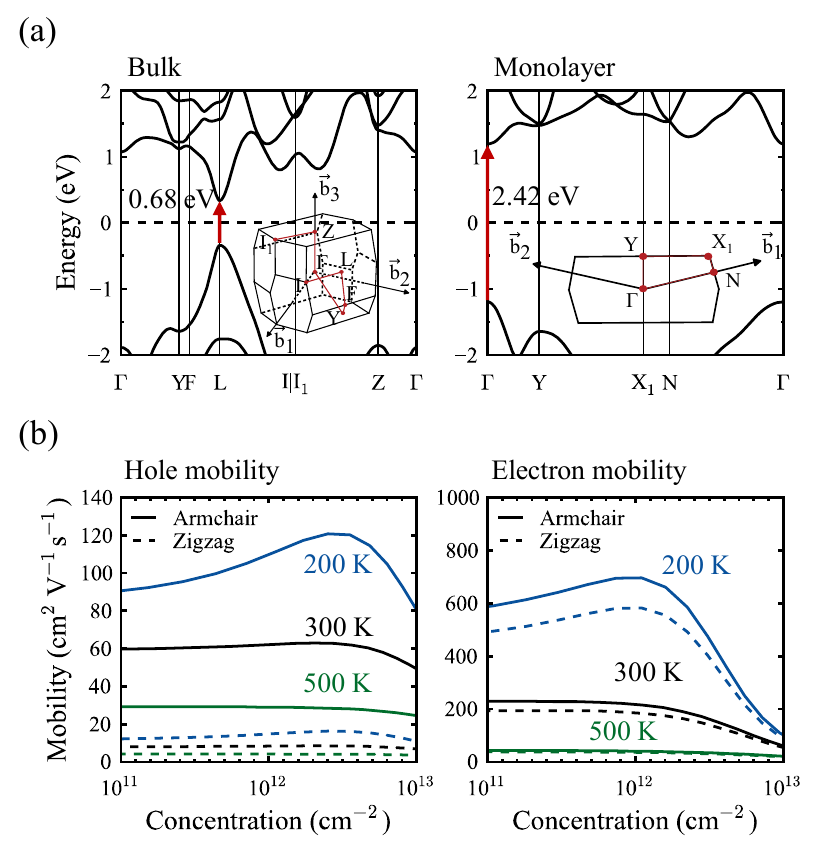}
\caption{\label{fig:3} (a) Electronic band structures of $\lambda$-P (green) in the bulk and monolayer structures using Wannier functions in the quasiparticle $G_{0}W_{0}$ calculations. 
(b) Hole and electron mobilities of green phosphorene as a function of carrier concentration along the armchair and zigzag directions. }
\end{figure}

Our results for the electronic structure of green phosphorus are shown in Fig. 3.
With the quasiparticle $G_{0}W_{0}$ approximation \cite{Hybertsen1986}, we obtained the direct band gap of 0.68 eV at the $L$ point in the BZ for the bulk structure.
The fundamental band gap is inversely proportional to the number of atomic layers,
without altering the direct band gap nature.
In the monolayer limit, we find the direct band gap of 2.42 eV at the BZ center.
This result indicates that green phosphorene is a semiconductor more suitable for electronic and optical applications, compared to blue phosphorene with the indirect band gap of 3.34 eV.
For green phosphorus in the $AA$ stacking, the band gap increases by about 0.4 eV and still varies inversely with the number of atomic layers.
Similar to black and blue phosphorene, the band gap of green phosphorene is sensitive to strain.
The gap size generally decreases under strain and the band gap nature changes from direct to indirect as strain changes from tensile to compressive \cite{*[{See Supplemental Material at }] [{}] supp}.

\begin{table}[!h]\centering
\begin{ruledtabular}
\caption{The band gaps ($E_{g}$) and exciton binding energies ($E_{ex}$) are compared for blue, black, and green phosphorene, based on quasiparticle $G_{0}W_{0}$, PBE, and hybrid functional (HSE06) calculations for $E_{g}$ and $G_{0}W_{0}$-BSE calculations for $E_{ex}$.
Here D and I denote the direct and indirect band gaps, respectively. }
\begin{tabular}{ccccc}
&  \multicolumn{3}{c}{$E_{g}$ (eV)} & \multicolumn{1}{c}{$E_{ex}$ (eV)}\\
\cline{2-4} \cline{5-5}
  & PBE & HSE06 & $G_{0}W_{0}$ & $G_{0}W_{0}$-BSE \\
\hline
Blue P	& 1.94 (I)	& 2.80 (I)	& 3.34 (I)	& 1.07  \\
Black P	& 0.90 (D)	& 1.60 (D)	& 1.91 (D)	& 0.71  \\
Green P	& 1.12 (D)	& 1.86 (D)	& 2.42 (D)	& 0.77  \\

\end{tabular}
\end{ruledtabular}
\end{table}

The optical absorption properties of green phosphorene are investigated by solving the Bethe-Salpeter equation (BSE) together with the quasiparticle $G_{0}W_{0}$ approximation \cite{Rohlfing2000}.
The inclusion of electron-hole interactions significantly changes the optical properties of green phosphorene.
In black phosphorene, the optical absorption spectra are known to be anisotropic, with the stronger exciton effect along the armchair direction \cite{Tran2014}, while blue phosphorene exhibits the isotropic absorption spectra.
Our $G_{0}W_{0}$-BSE calculations show the band gap of 1.91 eV and the first absorption peak at 1.20 eV (Table 2).
The exciton binding energy of 710 meV is in good agreement with the previous result of 800 meV \cite{Tran2014}.
Green phosphorene also shows the anisotropic absorption spectra, with the first peak at 1.65 eV along the armchair direction, and the exciton binding energy is 770 meV, similar to that of black phosphorene. 

Black phosphorene is known to have anisotropic effective masses for both hole and electron carriers, with the higher effective masses along the zigzag direction \cite{Wang2015}.
In green phosphorene, although the hole effective mass is reduced by about five times along the zigzag direction, it is less affected along the armchair direction, resulting in the strong anisotropy in the hole effective mass \cite{*[{See Supplemental Material at }] [{}] supp}.
On the other hand, since the electron effective mass is reduced by an order of magnitude along the zigzag direction, the nearly isotropic electron mobility is expected.

We calculated electron-phonon coupling matrices using the PBE functional and obtained scattering rates ($\omega^{el-ph}_{n,\vec{k}}$) and relaxation times ($\tau^{el-ph}_{n,\vec{k}}$) for the band $n$ and the wave vector $\vec{k}$, which are related to the imaginary part of electron self-energy ($\Im(\Sigma^{el-ph}_{n,\vec{k}})$) \cite{Giustino2007,Ponce2016},
\begin{align}
\omega^{el-ph}_{n,\vec{k}} = \frac{1}{\tau^{el-ph}_{n,\vec{k}}} = \frac{2}{\hbar}\Im(\Sigma^{el-ph}_{n,\vec{k}}).
\end{align}
Using the scattering rates in the semiclassical Boltzmann equation \cite{Pizzi2014}, the electron and hole mobilities were calculated as a function of carrier concentration [Fig. 3(b)].
As expected from the analysis of effective masses, green phosphorene exhibits the strong anisotropy in hole mobility, while the electron mobility is nearly isotropic above room temperature.
When temperature is lowered to 200 K, both the hole and electron mobilities increase owing to the suppression of phonon scattering.
For carrier concentrations up to $5 \times 10^{12}$ cm$^{-2}$, the electron mobility is calculated to be about 200 cm$^2$V$^{-1}$s$^{-1}$ at 300 K, which is much higher than the hole mobility of about 60 cm$^2$V$^{-1}$s$^{-1}$ along the armchair direction due to the lower scattering rates near the conduction band minimum \cite{*[{See Supplemental Material at }] [{}] supp}.
For black phosphorene, the hole and electron mobilities along the armchair direction are about 150 and 70 cm$^2$V$^{-1}$s$^{-1}$ at 300 K, respectively \cite{*[{See Supplemental Material at }] [{}] supp}.
While the calculated hole mobility is similar to the previous result of 170 cm$^2$V$^{-1}$s$^{-1}$ \cite{Liao2015}, it was experimentally reported that the hole mobility varies from about 200 to 50 cm$^2$V$^{-1}$s$^{-1}$ as the film thickness decreases to a few layers \cite{Liu2014}.
On the other hand, for blue phosphorene, we find the low isotropic mobilities of about 20 and 0.1 cm$^2$V$^{-1}$s$^{-1}$ for electron and hole carriers, respectively. 
Our results indicate that green phosphorene can serve as a potential material for $n$-type devices due to its high electron mobility.  

\begin{figure}[!h]
\includegraphics[width=\columnwidth]{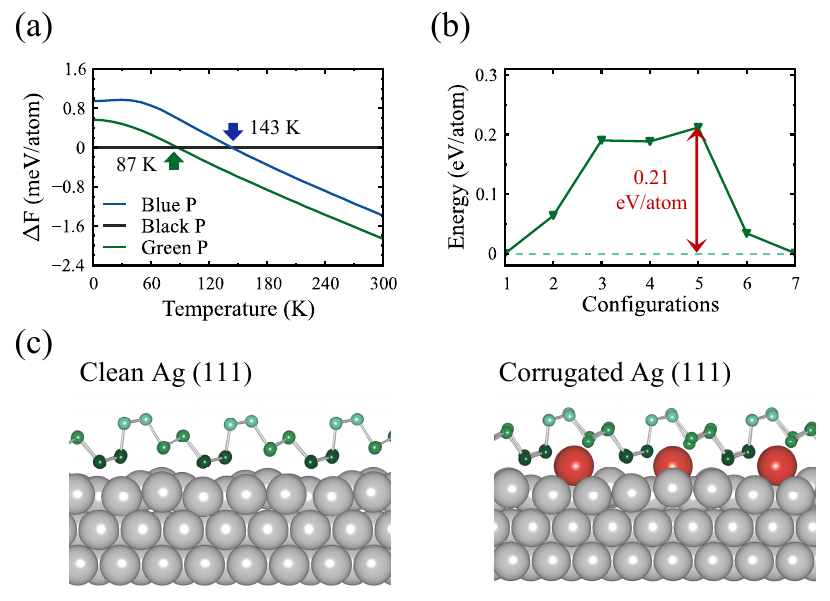}
\caption{\label{fig:4} (a) Helmholtz free energies ($\Delta F$) of blue and green phosphorene relative to black phosphorene as a function of temperature, with neglecting the thermal expansion effect.
(b) Total energy variation during the transformation from black to green phosphorene \cite{*[{See Supplemental Material at }] [{}] supp}.
(c) Atomic structures of green phosphorene on clean and corrugated Ag(111) surfaces.
Red balls represent the periodically corrugated Ag atoms.}
\end{figure}

The effect of temperature on the relative stability of black, blue, and green phosphorene was studied by calculating the Helmholtz free energy that includes both the vibrational entropy and zero-point energy.
Our calculations show that blue phosphorene becomes more stable than black phosphorene above 143 K, consistent with the previous result of 135 K, which was based on the Gibbs free energy \cite{Aierken2015}.
However, we note that blue phosphorene is always metastable with respect to green phosphorene for temperatures up to 300 K.
In fact, black phosphorene is more likely to transform to green phosphorene at the lower temperature of 87 K [Fig. 4(a)].

We considered a possible transition pathway from black to green phosphorene, which is accompanied with bond breaking and rebonding relaxations \cite{*[{See Supplemental Material at }] [{}] supp}.
Using the nudged elastic band method, we estimated the energy barrier to be 0.21 eV/atom [Fig. 4(b)], lower than the previously reported barrier of 0.47 eV/atom for the transition from black to blue phosphorene \cite{Zhu2014}.
We also examined the stability of the three allotropes on metal substrates including Al, Au, and Ag \cite{*[{See Supplemental Material at }] [{}] supp}.
Blue phosphorene is the most favorable allotrope on the clean (111) surfaces of Au and Ag, while black phosphorene is more stable on the Al substrate.
The energetics of blue phosphorene is in good agreement with the recent experimental realization on the Au substrate \cite{Zhang2016}.
On the other hand, green phosphorene is well matched to wrinkled surfaces due to its buckled structure, as shown in Fig. 4(c).
Thus, green phosphorene has the lowest formation energy on the corrugated (111) surfaces of Au and Ag \cite{*[{See Supplemental Material at }] [{}] supp}, indicating that this new allotrope can be realized on corrugated metal surfaces.

In summary, we have predicted a novel P allotrope called green phosphorus using the evolutionary crystal structure search method. 
While blue phosphorus has the indirect band gaps, green phosphorus exhibits the direct band gap nature, regardless of the number of atomic layers.
The band gap sizes are tunable in the visible range by varying the film thickness, lying in between black and blue phosphorus.
Due to the strong anisotropy in the optical and transport properties, a monolayer of green phosphorus, termed green phosphorene, is promising for novel 2D devices employing the anisotropic properties.
Furthermore, our calculations suggest that the transition from black to green phosphorene can occur at temperatures above 87 K, and the synthesis of green phosphorene is possible on corrugated metal substrates.


\begin{acknowledgments}
W.H. Han thanks Dr. Seoung-Hun Kang and Mr. Seungjun Lee for useful discussions on EPW. 
This work was supported by Samsung Science and Technology Foundation under Grant No. SSTFBA1401-08.
\end{acknowledgments}

\end{document}